\documentclass[fleqn,usenatbib]{mnras}
\usepackage{newtxtext,newtxmath}
\usepackage[T1]{fontenc}
\DeclareRobustCommand{\VAN}[3]{#2}
\let\VANthebibliography\thebibliography
\def\thebibliography{\DeclareRobustCommand{\VAN}[3]{##3}\VANthebibliography}

\usepackage{graphicx}	
\usepackage{amsmath}	

\newcommand{\Msun}{\mathrm{M_{\sun}}}
\newcommand{\kmps}{\mathrm{km\,s^{-1}}}

\title[The natal kick of H\,1705--250]{On the natal kick of the black hole X-ray binary H\,1705--250}

\author[C. Dashwood Brown et al.]{
Cordelia Dashwood Brown,
Poshak Gandhi,
and Yue Zhao
\\
School of Physics and Astronomy, University of Southampton, Southampton SO17 1BJ, UK\\
}

\date{Accepted 2023 October 5. Received 2023 October 5; in original form 2023 Jun 23}
\pubyear{2015}

\begin{document}
\label{firstpage}
\pagerange{\pageref{firstpage}--\pageref{lastpage}}
\maketitle

\begin{abstract}
When a compact object is formed, an impulse (kick) will be imparted to the system by the mass lost during the core-collapse supernova (SN). A number of other mechanisms may impart an additional kick on the system, although evidence for these natal kicks in black hole systems remains limited. Updated \textit{Gaia} astrometry has recently identified a number of high peculiar velocity (in excess of Galactic motion) compact objects. Here, we focus on the black hole low-mass X-ray binary H\,1705--250, which has a peculiar velocity $\upsilon_{\mathrm{pec}}\,=\,221^{+101}_{-108}\,\mathrm{km}\,\mathrm{s}^{-1}$. Using population synthesis to reconstruct its evolutionary history (assuming formation via isolated binary evolution within the Galactic plane), we constrain the properties of the progenitor and pre-SN orbit. The magnitude of a kick solely due to mass loss is found to be $\sim$\,30\,km\,s$^{-1}$, which cannot account for the high present-day peculiar motion. We therefore deduce that the black hole received an additional natal kick at formation, and place limits on its magnitude, finding it to be $\sim$\,295\,km\,s$^{-1}$ (minimum 90\,km\,s$^{-1}$). This furthers the argument that these kicks are not limited to neutron stars. 
\end{abstract}




\section{Introduction}
The majority of stellar mass black holes (BHs) detected in the electromagnetic spectrum exist in binaries, where they are actively accreting from a companion star, primarily emitting in the X-ray band. Despite a growing catalogue of these BH X-ray binaries (XRBs) with confirmed masses and well-constrained astrometry, it remains unclear how these objects form. The consensus is that BHs form from the gravitational collapse of a massive star, with a core mass $>\,8\,\mathrm{M}_{\odot}$. The BH may form directly after collapse; alternatively, if the supernova (SN) is not energetic enough to successfully eject all it's envelope, the BH may form through fallback, after existing briefly as a neutron star (NS) \citep[see][and references therein]{Repetto15,Mirabel17,Mapelli21}. 

The deaths of massive stars can impart additional acceleration to compact objects at the instant of SN, an impulse referred to as a natal kick. There remains significant uncertainty in the role of natal kicks in the evolution of BH systems. Previous literature has made a distinction between kicks associated with NS \citep[e.g.,][]{Gunn1970,Trimble71,Lyne82,Lyne94,Brandt95,Hobbs05,Verbunt17} versus BH \citep[e.g.,][]{Jonker04,Gualandris05,Repetto12,janka13a,Mandel15}, with the former being more extensively studied. Whilst there exists evidence for NS kicks with both high and low velocities, evidence of BHs receiving natal kicks is less clear. 

Understanding these natal kicks is of great importance in shaping SN mechanisms and energetics, as well as providing constraints on the number of compact objects that may be retained in globular or young stellar clusters \citep[e.g.,][]{Pfahl02a, Strader12, Giesler18}. The exact mechanisms leading to these kicks is unknown, but they may be the result of the recoil due to baryonic ejecta \citep{Blaauw61,Brandt95,Nelemans99} or anisotropy in gravitational attraction due to asymetrically ejected mass \citep{janka13,janka17}. Alternatively, the recoil momentum could be due to asymmetric neutrino emission, related to the hydrodynamic processes within the SN \citep{chugai84,dorofeev85,arras99,Lai04}. 

These kicks leave imprints on the kinematics of XRBs, and the high peculiar velocity ($\upsilon_{\mathrm{pec}}$) of many XRBs supports the hypothesis that non-negligible natal kicks occurred at the instant of SN \citep{Mirabel01,gandhi19,Atri20,Fortin23,ODoherty23}, although this is only a proxy for kick velocity.

Previous studies have considered whether BHs receive substantial natal kicks. \cite{Willems05} considered the case of GRO J1655--40, re-constructing its evolutionary history to determine that the natal kick velocity ($\upsilon_{\mathrm{NK}}$) was small (a few 10s\,km\,s$^{-1}$), and in contrast to the high velocities associated in NSs. The high velocity of XTE J1118+480 was discussed by \cite{Mirabel01}, and follow-up work suggested a lower limit of $\upsilon_{\mathrm{NK}}$ as either 80\,km\,s$^{-1}$ \citep{Fragos09} or 93\,--\,106\,km\,s$^{-1}$ \citep{Repetto15}. A similar analysis of Cyg X-1 by \cite{Wong12} constrained $\upsilon_{\mathrm{NK}}\,<$\,77\,km\,s$^{-1}$, and \cite{Kimball23} found MAXI J1305--704 to be consistent with a high $\upsilon_{\mathrm{NK}}\,>$\,70\,km\,s$^{-1}$.

One source of post-SN velocity is symmetric mass-loss \citep{Nelemans99}, where mass is symmetrically and instantly ejected from the compact progenitor by the SN, resulting in a recoil kick to the system. Considering this in tandem with Kepler's laws, one can derive an expression for the velocity imparted to the binary system: 

\begin{equation}
    \upsilon_{\mathrm{MLK}} = 213 \times \frac{m}{M_{\odot}}  \frac{\Delta M}{M_{\odot}} \left(\frac{P_{\mathrm{circ}}}{\mathrm{day}}\right)^{-\frac{1}{3}}  \left(\frac{M_{\rm tot}}{M_{\odot}}\right)^{-\frac{5}{3}}   \mathrm{km\,s}^{-1},
    \label{eq:mlk}
\end{equation}

\noindent
where $m$ and $M_{\rm tot}$ are the luminous companion and post-SN total system masses, $\Delta M$ is the mass lost from the compact object progenitor due to SN, and $P_{\mathrm{circ}}$ is the post-SN circularised orbital period.

Eq.\,\ref{eq:mlk} yields an estimate of the post-SN velocity of a system purely perturbed by symmetric mass loss, subsequently referred to as the mass-loss kick velocity ($\upsilon_{\mathrm{MLK}}$). These are conceptually identical to Blaauw \citeyear{Blaauw61} kicks We can examine the peculiar velocities of each system compared to the expected $\upsilon_{\mathrm{MLK}}$ to determine if this provides sufficient momentum, or if an additional kick (i.e. a natal kick related to the mechanisms outlined above) is necessary.

Here we present evidence for another BH-XRB with a large $\upsilon_{\mathrm{pec}}$, and test the argument that large kicks can be imparted to BHs at formation by tracing its evolutionary history. Sec \ref{sec:observations-analysis} describes the observations used in this work, Sec \ref{sec:simulations} presents the results of simulations, and Sec \ref{sec:discussions} provides relevant discussions.

\section{Observations \& Analysis}
\label{sec:observations-analysis}
Our target of interest was selected from a catalogue of 89 compact object binaries that includes updated and improved calculated peculiar velocities presented by \cite{Zhao23} (hereafter Zhao23). The dataset is a comprehensive collation of Galactic binaries hosting compact objects with reliable astrometry from either \textit{Gaia} or radio interferometric or timing observations (Zhao23 and references therein). The updated parallax ($\varpi$), and proper motion values provided by \textit{Gaia} are used to derive 3D $\upsilon_\mathrm{pec}$, following the formulation described by \cite{Reid08}. Orbits are integrated back in time, and the values of $\upsilon_\mathrm{pec}$ as the binary passes the Galactic plane are recorded, as a proxy for the peculiar velocity at birth ($\upsilon_{\mathrm{pec}}^{z=0}$) --- assuming that the binaries are born in the Galactic plane (for details, see Zhao23). The primary findings of Zhao23 are an anti-correlation between $\upsilon_{\mathrm{pec}}$ and $M_{{\rm tot}}$; here we focus on the origins of high-$\upsilon_{\mathrm{pec}}$ systems and the implications for natal kicks. 

High natal kick candidates can be identified as follows. For each binary the maximum velocity associated solely with mass-loss ($\upsilon_{\mathrm{MLK}}$) can be estimated and compared to the calculated $\upsilon_{\mathrm{pec}}$. From the virial theorem, it follows that a system that loses more than half its total mass due to the SN will be disrupted; therefore the present-day $M_{\rm tot}$ of the system provides an upper limit of $\Delta M$. We also assume that the present-day circular orbital period ($P_{\mathrm{orb}}$) is consistent with the $P_{\mathrm{circ}}$ (i.e. there has been no significant change in $P_{\mathrm{orb}}$ due to mass transfer, magnetic braking, etc.). This assumption is later improved upon with simulations. The $\upsilon_{\mathrm{MLK}}$ can be estimated from Eq.\,\ref{eq:mlk}.

\begin{figure}
 \includegraphics[width=8.5cm]{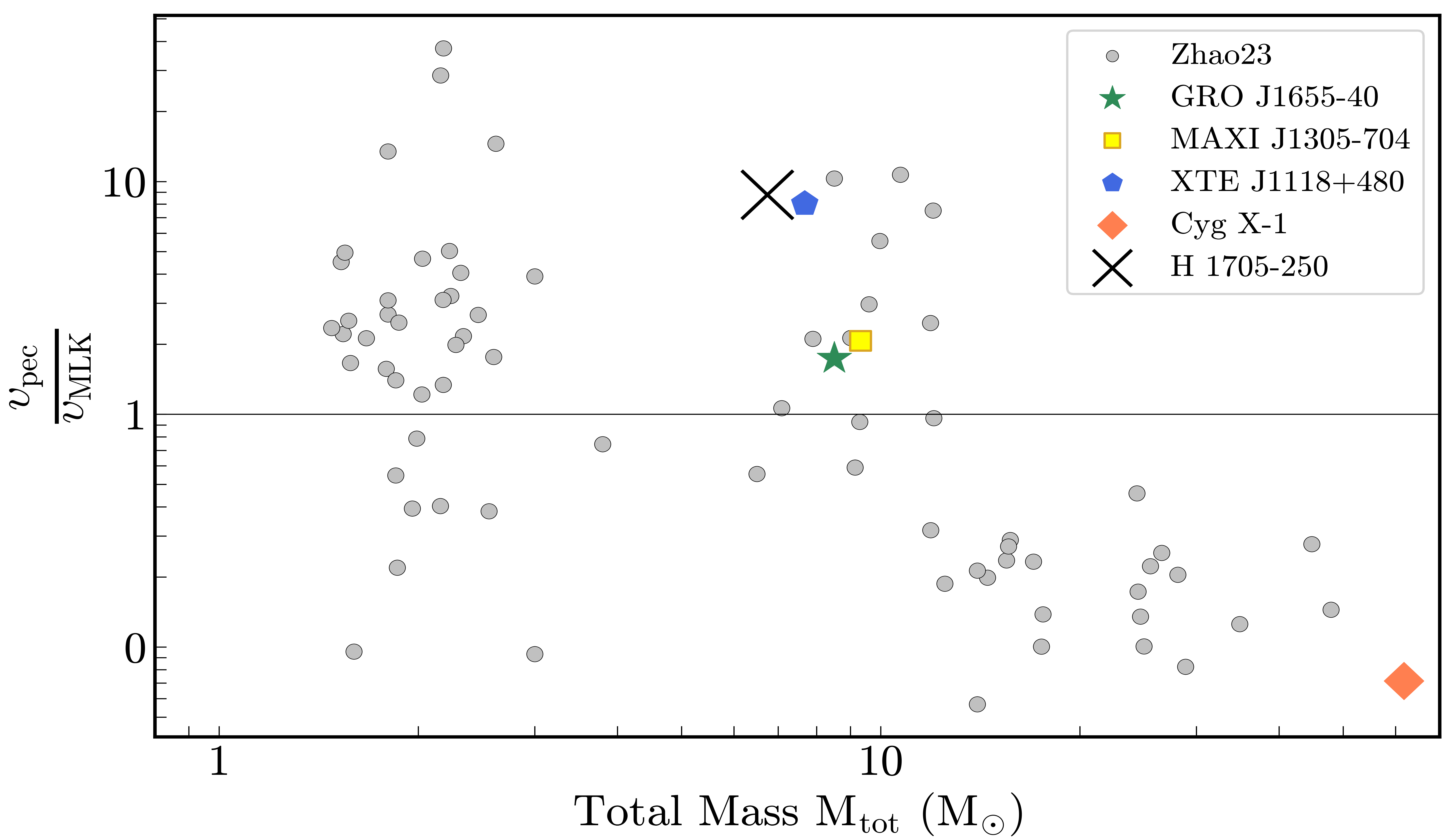}
\caption{Present-day peculiar velocity ($\upsilon_{\mathrm{pec}}$) compared with the estimated mass-loss kick velocity ($\upsilon_{\mathrm{MLK}}$) as a function of total system mass for the Zhao23 sample. H\,1705--250 (the subject of this paper) is highlighted by the black cross. Previous studies of black holes are distinguished by colour and shape: GRO J1655--40 \citep{Willems05}, XTE J1118+480 \citep{Fragos09}, Cyg X-1 \citep{Wong12}, and MAXI J1305--704 \citep{Kimball23}.}
\label{fig:1}
\end{figure}

Where systems have a $\upsilon_{\mathrm{pec}}$ higher than $\upsilon_{\mathrm{MLK}}$, it can be inferred that an additional source of momentum may have been imparted --- an additional kick related to mechanisms other than instantaneous mass ejection. Fig \ref{fig:1} shows the estimated $\upsilon_{\mathrm{MLK}}$ compared to the measured $\upsilon_{\mathrm{pec}}$ for the Zhao23 sample. We highlight systems which have been the subject of previous targeted natal kick studies and the focus of this paper (H\,1705--250).

There are a number of systems (both NS \& BH) which appear to require additional kicks to explain their $\upsilon_{\mathrm{pec}}$. That a large proportion of BH systems seem to require additional natal kicks is evidence that, whilst there may be differences in the kick origins and mechanisms, both types of systems can experience additional natal kicks at birth. 

Whilst the estimated $\upsilon_\mathrm{MLK}$ is a good litmus test for systems that may require additional kicks, it cannot be used in isolation, and detailed evolutionary histories need to be examined to infer more robust conclusions. We focus on one candidate that appears to have been kicked: H\,1705--250 (or Nova\,Oph\,1997, hereafter H1705), a short-period BH low mass X-ray binary (LMXB), with $P_\mathrm{orb} = 12.51 \pm 0.03$\,hrs, and masses $M_\mathrm{BH} = 6.4\pm 1.5\,\Msun$ and $m = 0.34\pm 0.08\,\Msun$  for the BH and stellar companion, respectively \citep{Harlaftis97,Remillard96}. This system has an observed space velocity $\upsilon_{\mathrm{pec}} = 221.2^{+100.8}_{-108.0}\,\kmps$ (Zhao23). We emphasise that these are not sigma errors, but rather the result of Monte-Carlo calculations of peculiar velocity using the astrometric and radial velocity errors. We chose to focus on this system as Zhao23 suggests higher mass systems are less likely to receive high natal kicks, and whilst the velocity of this source has been commented on in previous literature \citep{Mandel15,Repetto15}, there are inconsistencies in the peculiar velocities quoted in these papers. Based on the assumptions outlined above, $\upsilon_{\mathrm{MLK}}$ is estimated to be $25\pm 7\,\kmps$ (incorporating 1$\sigma$ uncertainties in the component masses and $P_{\mathrm{orb}}$). We carry out detailed simulations of this system to identify its evolutionary history, and the conditions required to account for the present-day characteristics.

\section{Simulations}
\label{sec:simulations}
In order to understand the history of H1705 and the impact of the SN, we generate a population of 6 million binaries and identify those that evolve to match the current system characteristics (primarily the masses and orbital period). We do this with \textsc{Compact Object Synthesis and Monte Carlo Investigation Code} (\textsc{cosmic}), a binary population synthesis code adapted from \textsc{Binary Stellar Evolution code (BSE)} to include modified and updated evolution prescriptions and parameters \citep{Breivik20}. 

We generate 6\,million binaries under a range of different initial conditions and track their evolution, searching for systems that will evolve to resemble H1705 following the first SN. The majority of evolutionary parameters and prescriptions are kept constant, with only the natal kick inputs changing, to ensure we generate systems with a wide distribution of natal kick velocities ($\upsilon_{\mathrm{NK}}$) to determine which of these is most favoured. Our pre-SN masses spanned 2\,--\,20\,$M_{\odot}$ and 0.1\,--\,2\,$M_{\odot}$ for the compact progenitor and stellar companion, respectively, and $P_{\mathrm{orb}}$ = 2\,--\,48\,hrs. Natal kick magnitudes were drawn from a log-uniform distribution where $10^{0.1} \leq\upsilon_{\mathrm{NK}} \leq 10^3\,\kmps$, such that none of the current natal kick models are favoured. Our aim is to identify the magnitude of $\upsilon_{\mathrm{NK}}$ needed to explain the present-day $\upsilon_\mathrm{pec}$, and therefore we chose this input distribution such that $\upsilon_{\mathrm{NK}}$ encompass a broad range of magnitudes.
Initial eccentricities for all systems were set at zero; this is a fair assumption given that tidal forces will act to circularise binaries quickly, particularly closely separated binaries such as these. Note that whilst evolutionary uncertainties such as common envelope efficiency, rates of mass-loss from stellar wind, and changes in $P_{\mathrm{orb}}$ associated with magnetic braking may all impact the evolution of simulated binaries, the focus is on the effect of kicks. 
The common envelope efficiency parameter $\alpha$ was drawn from a uniform distribution between 0.1 and 1, and stellar metallicity for both the compact progenitor and stellar companion is set to 0.02. Mass transfer is treated according to the standard BSE choice, outlined in \cite{Hurley02}, and we use the magnetic braking prescription from \cite{Ivanova03}. We are not proposing any one of these choices to be `correct' over alternative theories, but rather ensuring that the effects of kicks imparted on the compact objects are isolated from models involved in binary evolution.

\begin{figure}
\includegraphics[width=8.5cm]{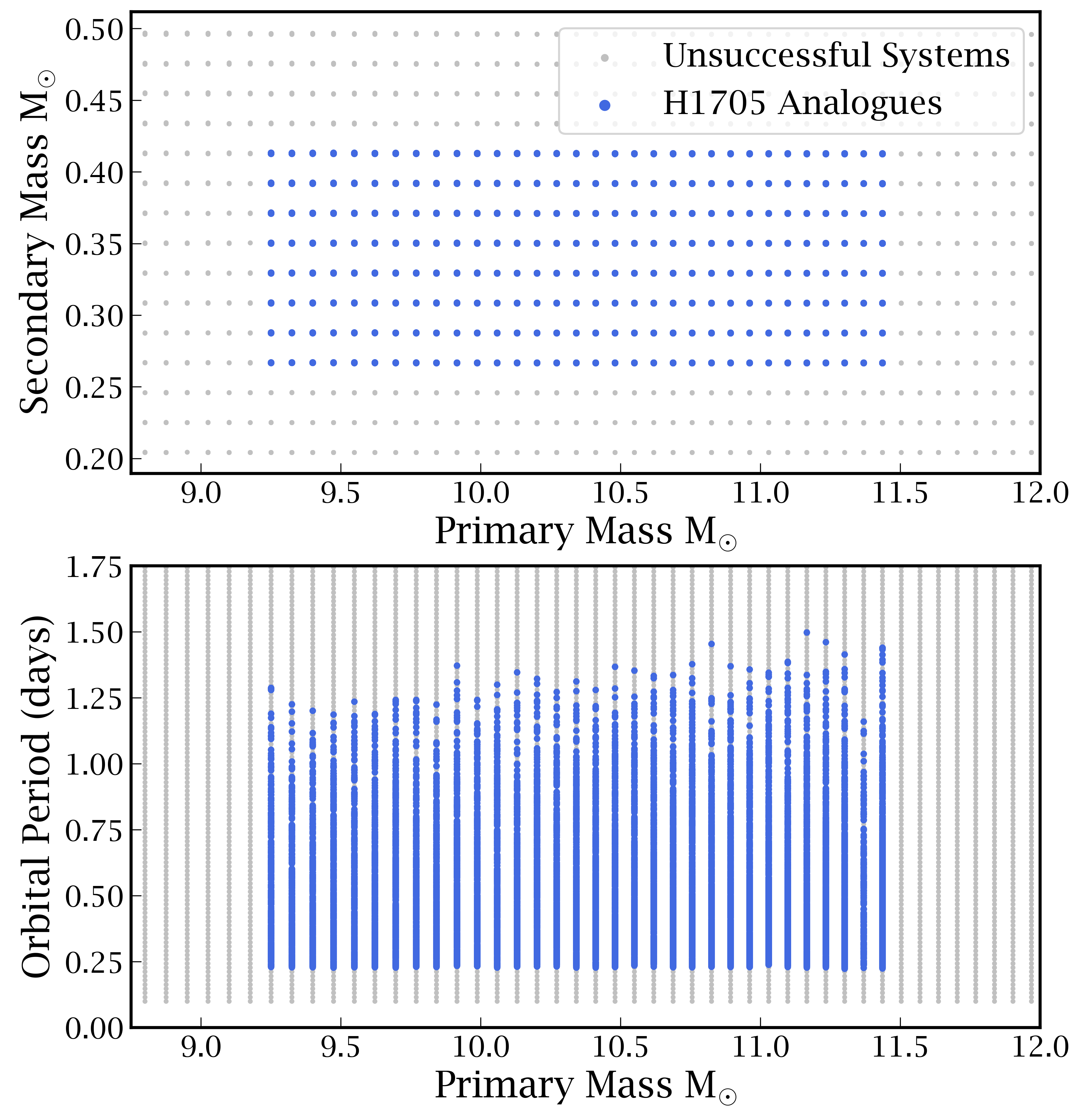}
\caption{Properties of H1705 progenitor systems, at the time of SN. Blue markers represent systems that evolve to form H1705 analogues after the SN, whilst grey systems are unsuccessful (either disrupted or merged during SN, or evolving to form different systems). Upper panel shows the masses of the different components (with the primary being the BH progenitor). Lower panel shows primary mass and orbital period}
\label{fig:2}
\end{figure}

The evolution of H1705 appears to follow `standard' binary evolution. The more massive component of the initial binary evolves, eventually breaching its Roche-lobe and resulting in mass transfer; it then undergoes a core-collapse supernova, leaving behind a BH. The results of these simulations are shown in Fig.\,\ref{fig:2}. There is a well-defined region of parameter space prior to the SN that is conducive to forming systems resembling H1705 (referred to hereon as H1705 analogues). At SN, the BH progenitor (a stripped Helium star) likely had a mass of 9.25 -- 11.5\,$\Msun$, and the companion had a mass of 0.25 -- 0.45\,$\Msun$. The expected $P_\mathrm{orb} \approx$ 0.2 -- 1.5\, days, with an orbital separation of 0.02 -- 0.06\,AU. 

Within our sample of 6\,million binaries, we focus on 2\,million where the pre-SN characteristics resemble those of the H1705 analogues. Of these, 0.5 million go on to form H1705 analogues, and the remainder act as a control sample to identify the effects of SN and natal kicks.

\section{Discussions}
\label{sec:discussions}
The components of our simulated systems have mean post-SN masses $M_{\mathrm{BH}}=6.49\pm 0.89\,\Msun$ and $m = 0.34\pm 0.05\,\Msun$, and $P_\mathrm{orb} = 12.4\pm 1.8$\,hrs, matching the observed system parameters well. Systems where the BH progenitor mass $\gtrapprox 11.5\,\Msun$ at the time of SN results in a post-SN $M_\mathrm{BH} > 9\,\Msun$. Conversely, if the primary mass $\lessapprox 8.5\,\Msun$, then systems are far more likely ($>80\%$) to be disrupted, and the surviving systems have $M_\mathrm{BH} \lessapprox 3.5\,\Msun$. Systems with pre-SN $P_\mathrm{orb} > 2\,$days are typically disrupted or forced into wider orbits by the SN. The BH progenitor in the H1705 analogues loses about 4\,$\Msun$: 30 -- 45\,\% of the $M_{\rm tot}$ at the time of SN.

\subsection{The effect of natal kicks on system survival}
Of the sample of binaries with pre-SN conditions matching those of H1705 analogues, 42\% will be disrupted by the SN, and a further 5\% will merge. Only $\approx$\,25\% form H1705 analogues. Systems with orbital periods $<$0.5\,days are more likely to merge than longer period systems, and all systems with $P_\mathrm{orb} < 0.24$\,days will merge. However, there is no obvious dependence on the masses. Figure \ref{fig:3} shows $\upsilon_{\mathrm{NK}}$ \& $\upsilon_{\mathrm{MLK}}$ for our simulated systems, distinguishing those disrupted by SN, those which do not evolve to match the characteristics of H1705, and H1705 analogues.
The majority ($\sim$\,70\%) of systems with $\upsilon_{\mathrm{NK}}$\,$<$\,50\,km\,s$^{-1}$ merge. Systems that are disrupted by the SN have, on average, higher $\upsilon_{\mathrm{NK}}$ than their surviving counterparts, and all systems where $\upsilon_{\mathrm{NK}}$\,>\,1000\,km\,s$^{-1}$ are disrupted.

\subsection{The effect of natal kicks on system velocity}

\begin{figure}
 \includegraphics[width=8.5cm]{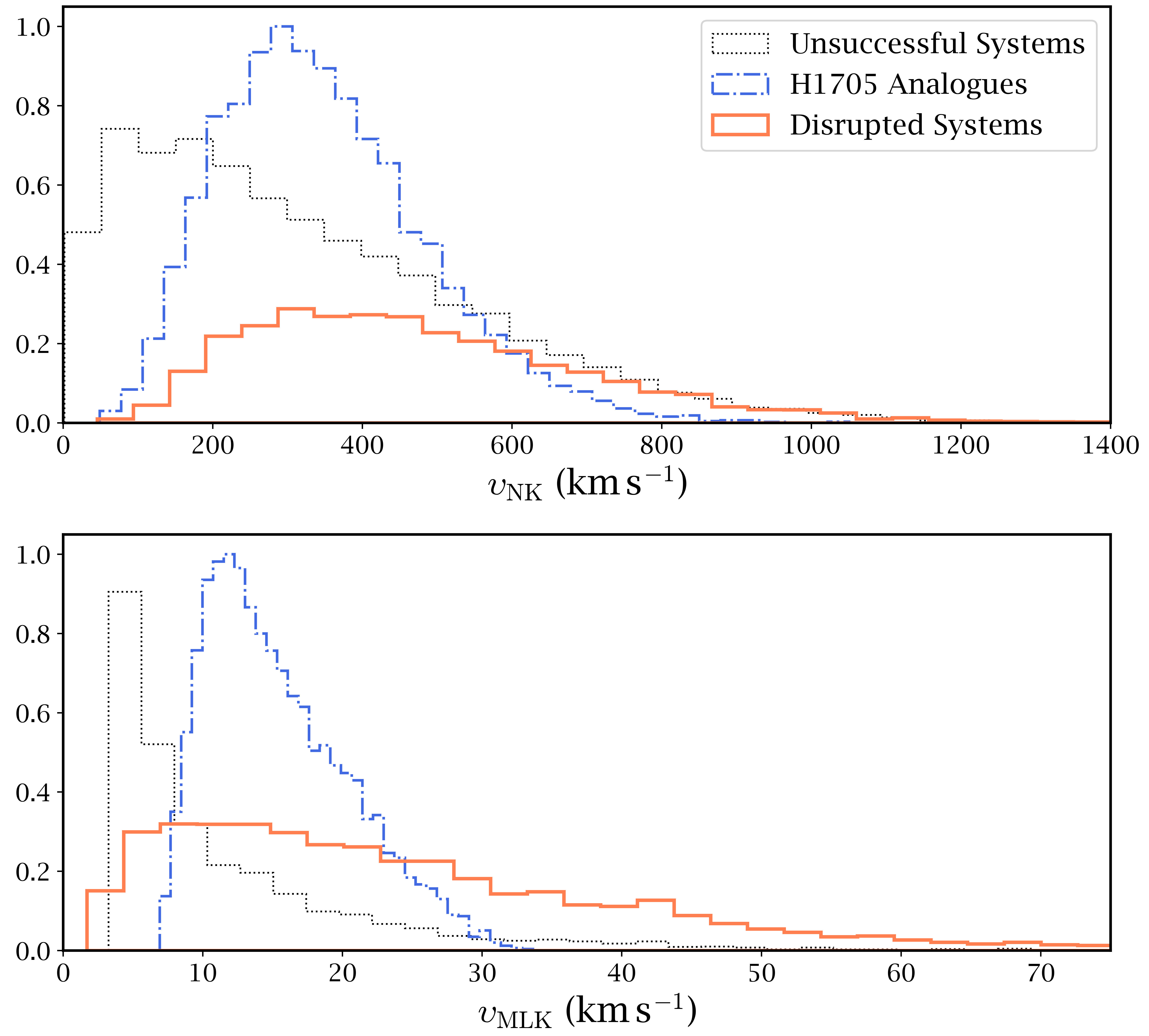}
\caption{Histograms showing the natal kick velocity ($\upsilon_{\mathrm{NK}}$) and mass-loss kick velocity ($\upsilon_{\mathrm{MLK}}$) of simulated systems. Blue (dashed line) systems mark H1705 analogues, grey (dotted line) marks those binaries which go on to evolve into systems distinct from H1705 (including systems that merge), and orange (solid line) marks systems which are disrupted during the SN.}
\label{fig:3}
\end{figure}

Simulations indicate that the compact progenitor loses $\sim$\,4\,M$_{\odot}$ as a consequence of the supernova - 40\% of its initial mass. Calculating the magnitude $\upsilon_{\mathrm{MLK}}$ for each system shows an average $\upsilon_{\mathrm{MLK}}\,\approx\,$16\,km\,s$^{-1}$ for H1705 analogues, with 95\% of systems having $\upsilon_{\mathrm{MLK}}\,<$\,25\,km\,s$^{-1}$. The distribution of $\upsilon_{\mathrm{MLK}}$ is shown in Fig. \ref{fig:3}. For the H1705 analogues, $\upsilon_{\mathrm{MLK}}$ does not exceed 35\,km\,s$^{-1}$; we can therefore infer that the momentum imparted to the surviving system from the mass ejection in the H1705 system cannot explain the high system velocity observed. An additional source of momentum, due to natal kick related to mechanisms other than mass ejection, is required.  

\begin{figure}
 \includegraphics[width=8.5cm]{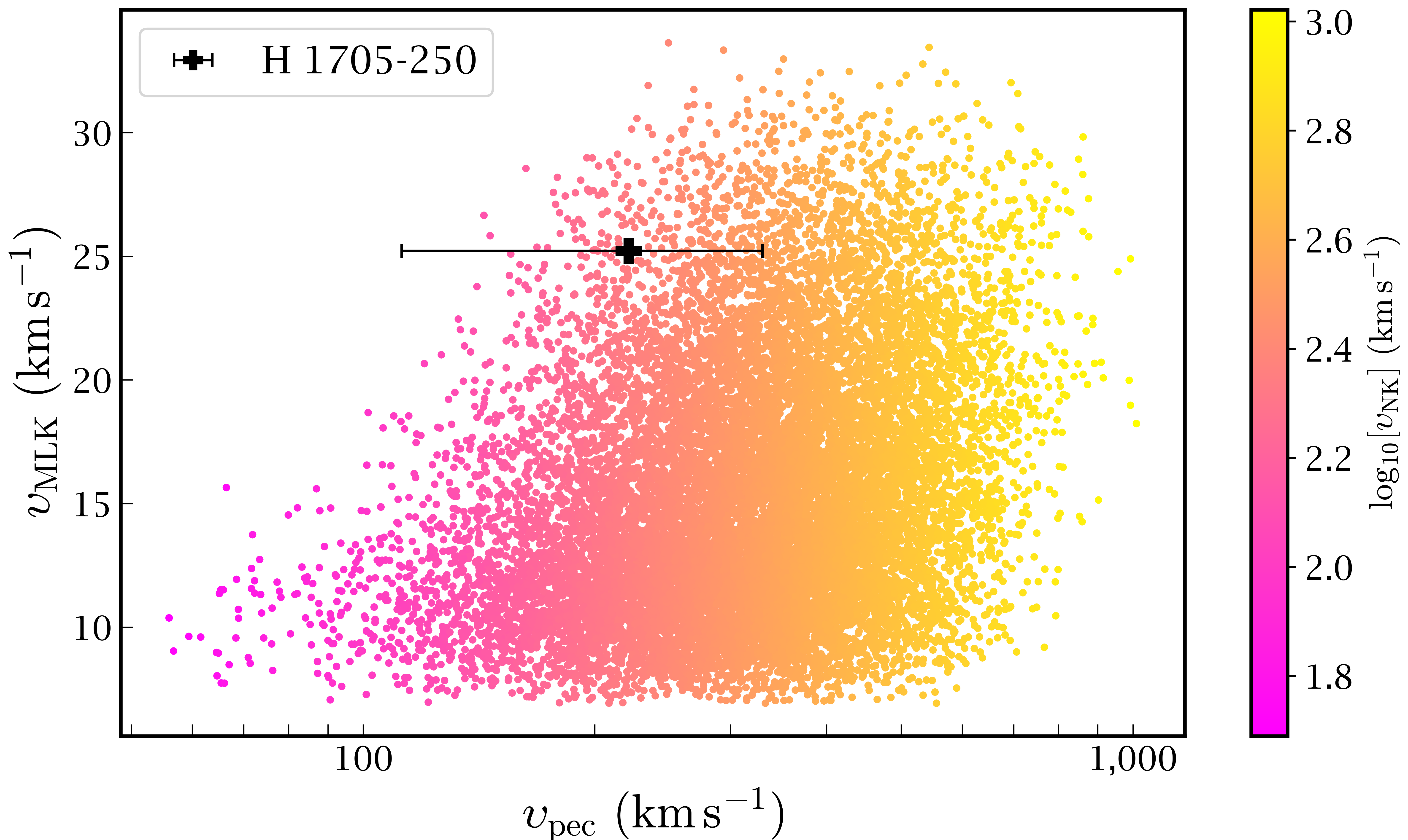}
\caption{Post-SN peculiar velocity ($\upsilon_{\mathrm{pec}}$) compared to the mass-loss kick velocity ($\upsilon_{\mathrm{MLK}}$) for all H1705 analogues. The colour bar indicates the velocity of the natal kick imparted on the system ($\log_{10}(\upsilon_{\mathrm{NK}})$). The observed present-day $\upsilon_{\mathrm{pec}}$ of H1705 is indicated by the black marker, with errors.}
\label{fig:4}
\end{figure}

In Fig. \ref{fig:4} we show the range of $\upsilon_{\mathrm{pec}}$ of our H1705 analogues compared to the $\upsilon_{\mathrm{MLK}}$, colour-coded by the magnitude of $\upsilon_\mathrm{NK}$. Around 53\% of systems have $\upsilon_\mathrm{pec}$ consistent with $\upsilon_{\mathrm{pec}}^{z=0}$ (Zhao23). Assuming this is an accurate estimate of the birth $\upsilon_\mathrm{pec}$, H1705 must have received an additional $\large\upsilon_{\mathrm{NK}}\,\gtrapprox\,97$\,km\,s$^{-1}$. For systems where $100 < \upsilon_\mathrm{pec}< 200\,\kmps$ (i.e., the lower end of estimates for $\upsilon_{\mathrm{pec}}$ within errors), $\upsilon_{\mathrm{NK}}$ must be in excess of $95\,\kmps$. These values are lower than those found by (\cite{Repetto15}, where $\upsilon_\mathrm{NK} \approx$ 361-- 441\,$\kmps$), which is due to the different methods in inferring the $\upsilon_{\mathrm{pec}}$ at formation --- however, the primary findings remain the same: the BH system H1705 must have received an additional kick in line with those expected for NS systems.

\subsection{Caveats}
\subsubsection{H1705 distance uncertainties}
\begin{figure}
 \includegraphics[width=8.5cm]{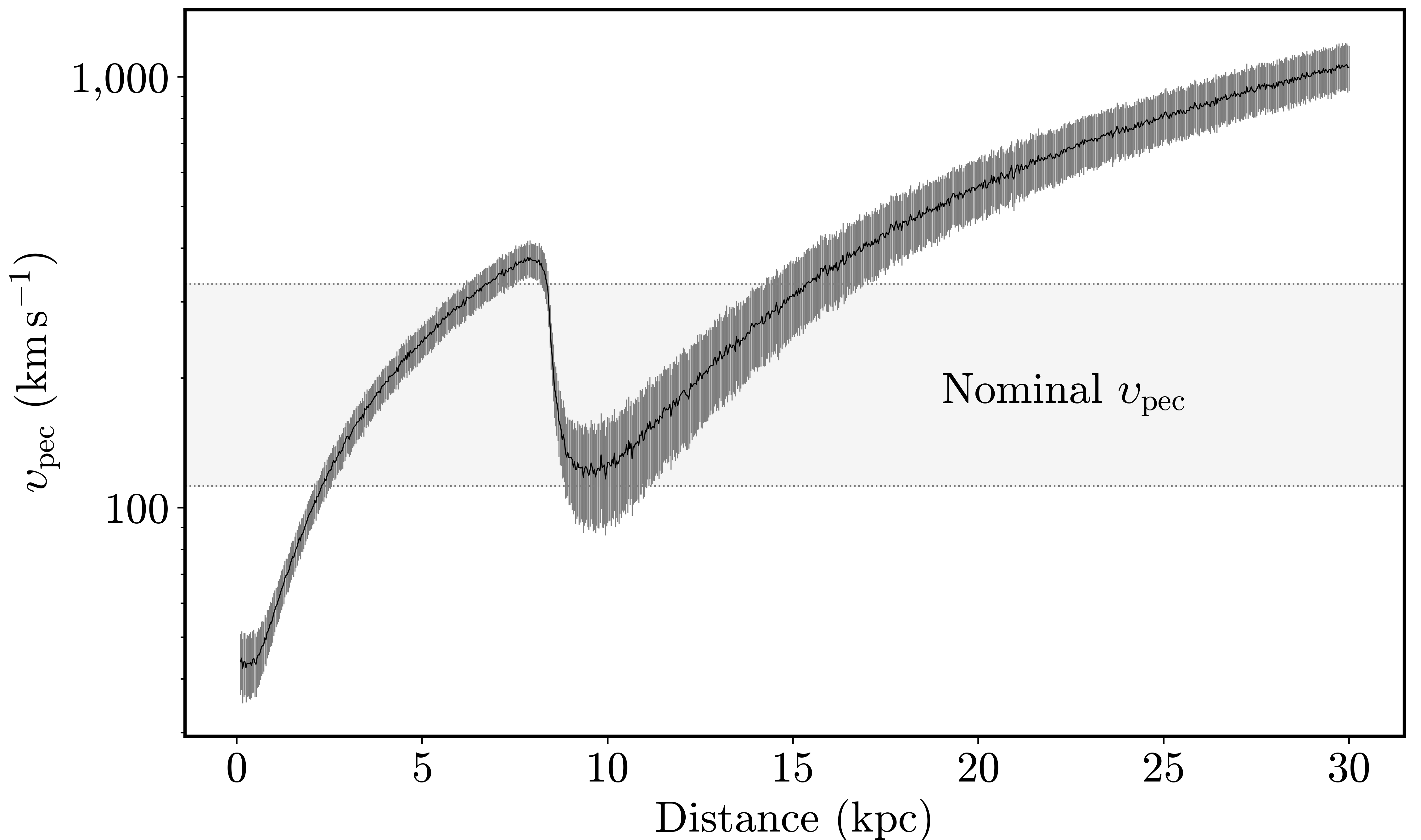}
 \caption{Peculiar velocity ($\upsilon_{\mathrm{pec}}$) as a function of distance to H1705. The errors (dark grey) are derived from the 1-$\sigma$ uncertainties in parallax, proper motions, and radial velocity. The light-shaded region indicates the quoted $\upsilon_{\mathrm{pec}}$ range ($221.2^{+100.8}_{-108.0}\,\mathrm{km}\,\mathrm{s}^{-1}$).}
\label{fig:5}
\end{figure}

The distance to H1705 is still highly uncertain. The \textit{Gaia} parallax measurement is only weakly significant, at $\varpi\,\approx 2.15\,\pm\,1.67\,mas$. Combined with exponential distance priors (Zhao23), the \textit{Gaia} predicted distance peaks at $\sim$\,3\,kpc, but with a significant tail to >10\,kpc. Future data releases should help narrow this prediction. \cite{Jonker04} determine the distance to H1705 to be 8.6\,$\pm$\,2.1\,kpc, and this is the value used by Zhao23 to calculate $\upsilon_{\mathrm{pec}}$. In Figure \ref{fig:5} we show the effect of varying distances in calculating the peculiar velocity. It is evident that for distances $>$\,2.5\,kpc, $\upsilon_{\mathrm{pec}}$ is substantial, at $>100$\,km\,s$^{-1}$. For this reason, we do not believe that the uncertainty in the distance to H1705 has a significant impact on our analysis.  

\subsubsection{Calculations of peculiar velocity}
One method of estimating peculiar velocity at birth (assuming birth in the Galactic plane) is by the velocity needed to reach its current location above the Galactic plane:
\begin{equation}
    \upsilon_{\perp,\mathrm{min}} = \sqrt{2\left [\Phi(R_{0},z) - \Phi(R_{0},0) \right]}
\end{equation}
where $\Phi(R,z)$ is the Galactic potential, and $R_{0}$ is the projected distance from the Galactic center \citep{Repetto12}. 

The uncertainty in distance means estimates of initial velocity from the above may be unreliable. This was highlighted by \cite{Mandel15}, who suggests that the uncertainties in distance led to an overestimation of the initial velocity, by up to an order of magnitude, and consequently, this source is therefore not suggestive of a significant natal kick. If one assumes the lowest literature distance, then the inferred velocity at birth (within the Galactic plane) would be less than 50\,km\,s$^{-1}$ and therefore explainable with a mass-loss kick alone (though higher kicks are not ruled out in general).

In this analysis, we consider the peculiar velocities calculated from the current positions and proper motions. \textit{Gaia} astrometry has placed solid constraints on the system's proper motion, despite $\varpi$ remaining poorly measured ($\frac{\varpi}{\sigma_{\varpi}}\,\sim$\,1.29). We comment that the uncertainties in distance and proper motions ($\mu_{\mathrm{RA}}\approx-6.97\pm2.83$\,mas\,yr$^{-1}$, $\mu_{\mathrm{DEC}}\approx-8.51\pm1.38$\,mas\,yr$^{-1}$) result in the high uncertainty of present-day $\upsilon_{\mathrm{pec}}$ ($221.2^{+100.8}_{-108.0}\,\mathrm{km}\,\mathrm{s}^{-1}$). However, even the lower limit on $\upsilon_{\mathrm{pec}}$ is significantly higher than both the estimated $\upsilon_{\mathrm{MLK}}$ and the mean peculiar velocities of all systems within the Zhao23 sample (we again note that the peculiar velocity distribution of H1705 is not Gaussian).

\subsubsection{System origin}
H1705 lies 1.35\,kpc above the Galactic plane, just above the Galactic bulge. Evolving the system back in time suggests that the system may have crossed the plane at some point in its history, and therefore could have formed here. However, given the height above the Galactic plane, we cannot exclude the possibility that H1705 instead formed within a globular cluster via dynamical interaction, and was subsequently ejected. In this scenario, some of our analysis would not be applicable. Where the velocity dispersion of the Galactic disc is $\sim$\,40\,km\,s$^{-1}$ \citep{Carlberg85}, the dispersion in the Galactic halo is larger: 50--120\,km\,s$^{-1}$ \citep{Battaglia05, Brown10}. However, the present-day $\upsilon_{\mathrm{pec}}$ of H1705 is still excessive compared to these estimates. The magnitude of the present-day $\upsilon_{\mathrm{pec}}$, neglecting any inference on the peculiar velocity at birth, is large enough that a high NK cannot immediately be ruled out if the system were to form in a GC. More detailed analysis and kinematic study of this system is needed to definitively determine the system's origin.

\subsubsection{Simulation restrictions}
As discussed in Sec \ref{sec:simulations}, the simulations carried out by \textsc{cosmic} do not explore the full range of parameters involved in binary evolution. For instance, our simulations are restricted to low eccentricities. Preliminary tests suggest that higher initial eccentricities do not preclude systems from evolving to resemble H1705, and $\upsilon_{\mathrm{NK}}$ must still be large in order to explain the present-day system kinematics. We also ran smaller-scale simulations whilst varying stellar metallicity by a factor of two each way, to determine if this has a significant effect on our results. Whilst the parameter space for systems that form H1705 analogues is slightly altered (e.g., doubling stellar metallicity results in, on average, slightly higher compact progenitor masses), our core findings still hold. Any variations in mass are small enough such that $\upsilon_{\mathrm{MLK}}$ remains largely unchanged, and the system still requires a substantial additional natal kick in order to reproduce the observed $\upsilon{\mathrm{pec}}$. Factors such as differing descriptions of mass-transfer, winds, common envelopes, magnetic braking, etc. were kept constant in all simulations. These factors are well-known uncertainties for binary population synthesis. We have assumed `canonical' estimates of these factors here, which can all be considered as physically plausible. Pathological or more complex scenarios cannot be ruled out, but are beyond the scope of this work.

\subsubsection{Kick mechanisms}
As discussed in Section 1, there are a range of theories regarding the origin of natal kicks. In this work, we do not constrain the nature of the mechanisms behind the kick imparted to H1705, nor suggest whether the kick is applied to the BH directly, or rather associated with a proto-NS before fallback. These questions can only be addressed by future detailed analysis of large samples and may allow us to infer likely sources of natal kicks and compare different populations.

It is also not fully understood how the momentum is imparted to the system and its constituents. It is believed that $\upsilon_{\mathrm{MLK}}$ is imparted to the center of mass of the system, whereas $\upsilon_{\mathrm{NK}}$, associated with mechanisms other than symmetric mass ejection, is imparted to the compact object. At this time there is no clear expression for how the system velocity will be affected by the combination of these two sources of momentum (and any others). In \textsc{cosmic}, the system velocity is derived such that it does not explicitly include the effects of mass-loss kicks as described by \cite{Nelemans99}. For this system, we have demonstrated the $\upsilon_{\mathrm{MLK}}$ is low, and therefore calculations carried out by \textsc{cosmic} are unlikely to significantly underestimate the $\upsilon_{\mathrm{pec}}$. However, it is important to recognise the limitations in our understanding of how the kinematics of a system post-SN evolve --- this will be the subject of future work.

\vspace{-0.4cm}
\section{Conclusions}
We have highlighted the high peculiar velocity of the BH-LMXB H\,1705--250, and presented evidence that the system requires a high natal kick (in addition to the imparted velocity associated with mass-loss due to supernova) to explain this. \textit{This marks another BH for which there is evidence of high natal kicks (i.e. comparable to the magnitude of NS kicks), implying that the discussion regarding natal kicks cannot be limited to NS}. We do not favour specific mechanisms or \textit{prescriptions} for natal kicks. Applying the same analysis used in this paper to a larger sample of binaries may help in constraining the nature of these kicks --- this will be the subject of future work. The birth site of H1705 is still degenerate; here we assume formation within the Galactic plane, however, we cannot rule out formation within the halo or a globular cluster. In this instance, the case for a high natal kick is less robust; though still appears to be required. 

\vspace{-0.4cm}
\section*{Acknowledgements}
We thank a referee for feedback and insight, and STFC for support. We also thank Christian Knigge for helpful comments and discussion. We thank Katie Breivik and all those involved in \textsc{cosmic} for their hard work, and for making the code publicly available.   

\vspace{-0.4cm}
\section*{Data Availability}
No new data are presented in this work. For details of compact object binaries (including H1705) refer to Zhao23. \textsc{cosmic} is publicly available on GitHub -- refer to \cite{Breivik20}. Future work will include the code used to track binary evolution with COSMIC.

\vspace{-0.4cm}
\bibliographystyle{mnras}
\bibliography{bib} 

\begin{thebibliography}{}
\makeatletter
\relax
\def\mn@urlcharsother{\let\do\@makeother \do\$\do\&\do\#\do\^\do\_\do\%\do\~}
\def\mn@doi{\begingroup\mn@urlcharsother \@ifnextchar [ {\mn@doi@}
  {\mn@doi@[]}}
\def\mn@doi@[#1]#2{\def\@tempa{#1}\ifx\@tempa\@empty \href
  {http://dx.doi.org/#2} {doi:#2}\else \href {http://dx.doi.org/#2} {#1}\fi
  \endgroup}
\def\mn@eprint#1#2{\mn@eprint@#1:#2::\@nil}
\def\mn@eprint@arXiv#1{\href {http://arxiv.org/abs/#1} {{\tt arXiv:#1}}}
\def\mn@eprint@dblp#1{\href {http://dblp.uni-trier.de/rec/bibtex/#1.xml}
  {dblp:#1}}
\def\mn@eprint@#1:#2:#3:#4\@nil{\def\@tempa {#1}\def\@tempb {#2}\def\@tempc
  {#3}\ifx \@tempc \@empty \let \@tempc \@tempb \let \@tempb \@tempa \fi \ifx
  \@tempb \@empty \def\@tempb {arXiv}\fi \@ifundefined
  {mn@eprint@\@tempb}{\@tempb:\@tempc}{\expandafter \expandafter \csname
  mn@eprint@\@tempb\endcsname \expandafter{\@tempc}}}

\bibitem[\protect\citeauthoryear{{Arras} \& {Lai}}{{Arras} \&
  {Lai}}{1999}]{arras99}
{Arras} P.,  {Lai} D.,  1999, \mn@doi [\apj] {10.1086/307407}, \href
  {https://ui.adsabs.harvard.edu/abs/1999ApJ...519..745A} {519, 745}

\bibitem[\protect\citeauthoryear{Atri et~al.,}{Atri et~al.}{2020}]{Atri20}
Atri P.,  et~al., 2020, \mn@doi [\mnras] {10.1093/mnrasl/slaa010}, 493, L81

\bibitem[\protect\citeauthoryear{{Battaglia} et~al.,}{{Battaglia}
  et~al.}{2005}]{Battaglia05}
{Battaglia} G.,  et~al., 2005, \mn@doi [\mnras]
  {10.1111/j.1365-2966.2005.09367.x}, \href
  {https://ui.adsabs.harvard.edu/abs/2005MNRAS.364..433B} {364, 433}

\bibitem[\protect\citeauthoryear{{Blaauw}}{{Blaauw}}{1961}]{Blaauw61}
{Blaauw} A.,  1961, Bulletin of the Astronomical Institutes of the Netherlands,
  \href {https://ui.adsabs.harvard.edu/abs/1961BAN....15..265B} {15, 265}

\bibitem[\protect\citeauthoryear{Brandt \& Podsiadlowski}{Brandt \&
  Podsiadlowski}{1995}]{Brandt95}
Brandt N.,  Podsiadlowski P.,  1995, \mn@doi [\mnras]
  {10.1093/mnras/274.2.461}, 274, 461–484

\bibitem[\protect\citeauthoryear{Breivik et~al.,}{Breivik
  et~al.}{2020}]{Breivik20}
Breivik K.,  et~al., 2020, \mn@doi [\apj] {10.3847/1538-4357/ab9d85}, 898, 71

\bibitem[\protect\citeauthoryear{{Brown}, {Geller}, {Kenyon}  \&
  {Diaferio}}{{Brown} et~al.}{2010}]{Brown10}
{Brown} W.~R.,  {Geller} M.~J.,  {Kenyon} S.~J.,   {Diaferio} A.,  2010,
  \mn@doi [\aj] {10.1088/0004-6256/139/1/59}, \href
  {https://ui.adsabs.harvard.edu/abs/2010AJ....139...59B} {139, 59}

\bibitem[\protect\citeauthoryear{{Carlberg}, {Dawson}, {Hsu}  \&
  {Vandenberg}}{{Carlberg} et~al.}{1985}]{Carlberg85}
{Carlberg} R.~G.,  {Dawson} P.~C.,  {Hsu} T.,   {Vandenberg} D.~A.,  1985,
  \mn@doi [\apj] {10.1086/163337}, \href
  {https://ui.adsabs.harvard.edu/abs/1985ApJ...294..674C} {294, 674}

\bibitem[\protect\citeauthoryear{{Chugai}}{{Chugai}}{1984}]{chugai84}
{Chugai} N.~N.,  1984, Soviet Astronomy Letters, \href
  {https://ui.adsabs.harvard.edu/abs/1984SvAL...10...87C} {10, 87}

\bibitem[\protect\citeauthoryear{{Dorofeev}, {Rodionov}  \&
  {Ternov}}{{Dorofeev} et~al.}{1985}]{dorofeev85}
{Dorofeev} O.~F.,  {Rodionov} V.~N.,   {Ternov} I.~M.,  1985, Soviet Astronomy
  Letters, \href {https://ui.adsabs.harvard.edu/abs/1985SvAL...11..123D} {11,
  123}

\bibitem[\protect\citeauthoryear{{Fortin}, {Garc{\'\i}a}, {Simaz Bunzel}  \&
  {Chaty}}{{Fortin} et~al.}{2023}]{Fortin23}
{Fortin} F.,  {Garc{\'\i}a} F.,  {Simaz Bunzel} A.,   {Chaty} S.,  2023,
  \mn@doi [\aap] {10.1051/0004-6361/202245236}, \href
  {https://ui.adsabs.harvard.edu/abs/2023A&A...671A.149F} {671, A149}

\bibitem[\protect\citeauthoryear{Fragos, Willems, Kalogera, Ivanova,
  Rockefeller, Fryer  \& Young}{Fragos et~al.}{2009}]{Fragos09}
Fragos T.,  Willems B.,  Kalogera V.,  Ivanova N.,  Rockefeller G.,  Fryer
  C.~L.,   Young P.~A.,  2009, \mn@doi [\apj] {10.1088/0004-637x/697/2/1057},
  697, 1057–1070

\bibitem[\protect\citeauthoryear{Gandhi, Rao, Johnson, Paice  \&
  Maccarone}{Gandhi et~al.}{2019}]{gandhi19}
Gandhi P.,  Rao A.,  Johnson M. A.~C.,  Paice J.~A.,   Maccarone T.~J.,  2019,
  \mn@doi [\mnras] {10.1093/\mnras/stz438}, 485, 2642

\bibitem[\protect\citeauthoryear{{Giesler}, {Clausen}  \& {Ott}}{{Giesler}
  et~al.}{2018}]{Giesler18}
{Giesler} M.,  {Clausen} D.,   {Ott} C.~D.,  2018, \mn@doi [\mnras]
  {10.1093/mnras/sty659}, \href
  {https://ui.adsabs.harvard.edu/abs/2018MNRAS.477.1853G} {477, 1853}

\bibitem[\protect\citeauthoryear{{Gualandris}, {Colpi}, {Portegies Zwart}  \&
  {Possenti}}{{Gualandris} et~al.}{2005}]{Gualandris05}
{Gualandris} A.,  {Colpi} M.,  {Portegies Zwart} S.,   {Possenti} A.,  2005,
  \mn@doi [\apj] {10.1086/426126}, \href
  {https://ui.adsabs.harvard.edu/abs/2005ApJ...618..845G} {618, 845}

\bibitem[\protect\citeauthoryear{{Gunn} \& {Ostriker}}{{Gunn} \&
  {Ostriker}}{1970}]{Gunn1970}
{Gunn} J.~E.,  {Ostriker} J.~P.,  1970, \mn@doi [\apj] {10.1086/150487}, \href
  {https://ui.adsabs.harvard.edu/abs/1970ApJ...160..979G} {160, 979}

\bibitem[\protect\citeauthoryear{{Harlaftis}, {Steeghs}, {Horne}  \&
  {Filippenko}}{{Harlaftis} et~al.}{1997}]{Harlaftis97}
{Harlaftis} E.~T.,  {Steeghs} D.,  {Horne} K.,   {Filippenko} A.~V.,  1997,
  \mn@doi [\aj] {10.1086/118548}, \href
  {https://ui.adsabs.harvard.edu/abs/1997AJ....114.1170H} {114, 1170}

\bibitem[\protect\citeauthoryear{{Hobbs}, {Lorimer}, {Lyne}  \&
  {Kramer}}{{Hobbs} et~al.}{2005}]{Hobbs05}
{Hobbs} G.,  {Lorimer} D.~R.,  {Lyne} A.~G.,   {Kramer} M.,  2005, \mn@doi
  [\mnras] {10.1111/j.1365-2966.2005.09087.x}, \href
  {https://ui.adsabs.harvard.edu/abs/2005MNRAS.360..974H} {360, 974}

\bibitem[\protect\citeauthoryear{{Hurley}, {Tout}  \& {Pols}}{{Hurley}
  et~al.}{2002}]{Hurley02}
{Hurley} J.~R.,  {Tout} C.~A.,   {Pols} O.~R.,  2002, \mn@doi [\mnras]
  {10.1046/j.1365-8711.2002.05038.x}, \href
  {https://ui.adsabs.harvard.edu/abs/2002MNRAS.329..897H} {329, 897}

\bibitem[\protect\citeauthoryear{{Ivanova} \& {Taam}}{{Ivanova} \&
  {Taam}}{2003}]{Ivanova03}
{Ivanova} N.,  {Taam} R.~E.,  2003, \mn@doi [\apj] {10.1086/379192}, \href
  {https://ui.adsabs.harvard.edu/abs/2003ApJ...599..516I} {599, 516}

\bibitem[\protect\citeauthoryear{{Janka}}{{Janka}}{2013a}]{janka13a}
{Janka} H.-T.,  2013a, \mn@doi [\mnras] {10.1093/mnras/stt1106}, \href
  {https://ui.adsabs.harvard.edu/abs/2013MNRAS.434.1355J} {434, 1355}

\bibitem[\protect\citeauthoryear{{Janka}}{{Janka}}{2013b}]{janka13}
{Janka} H.-T.,  2013b, \mn@doi [\mnras] {10.1093/\mnras/stt1106}, \href
  {https://ui.adsabs.harvard.edu/abs/2013\mnras.434.1355J} {434, 1355}

\bibitem[\protect\citeauthoryear{{Janka}}{{Janka}}{2017}]{janka17}
{Janka} H.-T.,  2017, \mn@doi [\apj] {10.3847/1538-4357/aa618e}, \href
  {https://ui.adsabs.harvard.edu/abs/2017ApJ...837...84J} {837, 84}

\bibitem[\protect\citeauthoryear{Jonker \& Nelemans}{Jonker \&
  Nelemans}{2004}]{Jonker04}
Jonker P.~G.,  Nelemans G.,  2004, \mn@doi [\mnras]
  {10.1111/j.1365-2966.2004.08193.x}, 354, 355–366

\bibitem[\protect\citeauthoryear{{Kimball} et~al.,}{{Kimball}
  et~al.}{2023}]{Kimball23}
{Kimball} C.,  et~al., 2023, \mn@doi [\apjl] {10.3847/2041-8213/ace526}, \href
  {https://ui.adsabs.harvard.edu/abs/2023ApJ...952L..34K} {952, L34}

\bibitem[\protect\citeauthoryear{{Lai}}{{Lai}}{2004}]{Lai04}
{Lai} D.,  2004, in {H{\"o}flich} P.,  {Kumar} P.,   {Wheeler} J.~C.,  eds,
  Cosmic explosions in three dimensions. p.~276 (\mn@eprint {arXiv}
  {astro-ph/0312542}), \mn@doi{10.48550/arXiv.astro-ph/0312542}

\bibitem[\protect\citeauthoryear{{Lyne} \& {Lorimer}}{{Lyne} \&
  {Lorimer}}{1994}]{Lyne94}
{Lyne} A.~G.,  {Lorimer} D.~R.,  1994, \mn@doi [Nature] {10.1038/369127a0},
  \href {https://ui.adsabs.harvard.edu/abs/1994Natur.369..127L} {369, 127}

\bibitem[\protect\citeauthoryear{Lyne, Anderson  \& Salter}{Lyne
  et~al.}{1982}]{Lyne82}
Lyne A.~G.,  Anderson B.,   Salter M.~J.,  1982, \mn@doi [\mnras]
  {10.1093/mnras/201.3.503}, 201, 503–520

\bibitem[\protect\citeauthoryear{Mandel}{Mandel}{2015}]{Mandel15}
Mandel I.,  2015, \mn@doi [\mnras] {10.1093/mnras/stv2733}, 456, 578–581

\bibitem[\protect\citeauthoryear{{Mapelli}}{{Mapelli}}{2021}]{Mapelli21}
{Mapelli} M.,  2021, in , Handbook of Gravitational Wave Astronomy.
p.~16, \mn@doi{10.1007/978-981-15-4702-7_16-1}

\bibitem[\protect\citeauthoryear{{Mirabel}}{{Mirabel}}{2017}]{Mirabel17}
{Mirabel} F.,  2017, \mn@doi [\nar] {10.1016/j.newar.2017.04.002}, \href
  {https://ui.adsabs.harvard.edu/abs/2017NewAR..78....1M} {78, 1}

\bibitem[\protect\citeauthoryear{{Mirabel}, {Dhawan}, {Mignani}, {Rodrigues}
  \& {Guglielmetti}}{{Mirabel} et~al.}{2001}]{Mirabel01}
{Mirabel} I.~F.,  {Dhawan} V.,  {Mignani} R.~P.,  {Rodrigues} I.,
  {Guglielmetti} F.,  2001, \mn@doi [\nat] {10.1038/35093060}, \href
  {https://ui.adsabs.harvard.edu/abs/2001Natur.413..139M} {413, 139}

\bibitem[\protect\citeauthoryear{Nelemans, Tauris  \& Heuvel}{Nelemans
  et~al.}{1999}]{Nelemans99}
Nelemans G.,  Tauris T.~M.,   Heuvel E.~P.,  1999, \mn@doi [Black Holes in
  Binaries and Galactic Nuclei: Diagnostics, Demography and Formation]
  {10.1007/10720995_68}, p. 312–313

\bibitem[\protect\citeauthoryear{{O'Doherty}, {Bahramian}, {Miller-Jones},
  {Goodwin}, {Mandel}, {Willcox}, {Atri}  \& {Strader}}{{O'Doherty}
  et~al.}{2023}]{ODoherty23}
{O'Doherty} T.~N.,  {Bahramian} A.,  {Miller-Jones} J. C.~A.,  {Goodwin} A.~J.,
   {Mandel} I.,  {Willcox} R.,  {Atri} P.,   {Strader} J.,  2023, \mn@doi
  [\mnras] {10.1093/mnras/stad680}, \href
  {https://ui.adsabs.harvard.edu/abs/2023MNRAS.521.2504O} {521, 2504}

\bibitem[\protect\citeauthoryear{{Pfahl}, {Rappaport}  \&
  {Podsiadlowski}}{{Pfahl} et~al.}{2002}]{Pfahl02a}
{Pfahl} E.,  {Rappaport} S.,   {Podsiadlowski} P.,  2002, \mn@doi [\apj]
  {10.1086/340494}, \href
  {https://ui.adsabs.harvard.edu/abs/2002ApJ...573..283P} {573, 283}

\bibitem[\protect\citeauthoryear{Reid et~al.,}{Reid et~al.}{2009}]{Reid08}
Reid M.~J.,  et~al., 2009, \mn@doi [\apj] {10.1088/0004-637x/700/1/137}, 700,
  137–148

\bibitem[\protect\citeauthoryear{Remillard, Orosz, McClintock  \&
  Bailyn}{Remillard et~al.}{1996}]{Remillard96}
Remillard R.~A.,  Orosz J.~A.,  McClintock J.~E.,   Bailyn C.~D.,  1996,
  \mn@doi [\apj] {10.1086/176885}, 459, 226

\bibitem[\protect\citeauthoryear{Repetto \& Nelemans}{Repetto \&
  Nelemans}{2015}]{Repetto15}
Repetto S.,  Nelemans G.,  2015, \mn@doi [\mnras] {10.1093/mnras/stv1753}, 453,
  3342–3356

\bibitem[\protect\citeauthoryear{Repetto, Davies  \& Sigurdsson}{Repetto
  et~al.}{2012}]{Repetto12}
Repetto S.,  Davies M.~B.,   Sigurdsson S.,  2012, \mn@doi [\mnras]
  {10.1111/j.1365-2966.2012.21549.x}, 425, 2799–2809

\bibitem[\protect\citeauthoryear{{Strader}, {Chomiuk}, {Maccarone},
  {Miller-Jones}  \& {Seth}}{{Strader} et~al.}{2012}]{Strader12}
{Strader} J.,  {Chomiuk} L.,  {Maccarone} T.~J.,  {Miller-Jones} J. C.~A.,
  {Seth} A.~C.,  2012, \mn@doi [\nat] {10.1038/nature11490}, \href
  {https://ui.adsabs.harvard.edu/abs/2012Natur.490...71S} {490, 71}

\bibitem[\protect\citeauthoryear{Trimble}{Trimble}{1971}]{Trimble71}
Trimble V.,  1971, \mn@doi [The Crab Nebula] {10.1007/978-94-010-3087-8_2}, p.
  12–21

\bibitem[\protect\citeauthoryear{{Verbunt}, {Igoshev}  \& {Cator}}{{Verbunt}
  et~al.}{2017}]{Verbunt17}
{Verbunt} F.,  {Igoshev} A.,   {Cator} E.,  2017, \mn@doi [\aap]
  {10.1051/0004-6361/201731518}, \href
  {https://ui.adsabs.harvard.edu/abs/2017A&A...608A..57V} {608, A57}

\bibitem[\protect\citeauthoryear{Willems, Henninger, Levin, Ivanova, Kalogera,
  McGhee, Timmes  \& Fryer}{Willems et~al.}{2005}]{Willems05}
Willems B.,  Henninger M.,  Levin T.,  Ivanova N.,  Kalogera V.,  McGhee K.,
  Timmes F.~X.,   Fryer C.~L.,  2005, \mn@doi [\apj] {10.1086/429557}, 625,
  324–346

\bibitem[\protect\citeauthoryear{{Wong}, {Valsecchi}, {Fragos}  \&
  {Kalogera}}{{Wong} et~al.}{2012}]{Wong12}
{Wong} T.-W.,  {Valsecchi} F.,  {Fragos} T.,   {Kalogera} V.,  2012, \mn@doi
  [\apj] {10.1088/0004-637X/747/2/111}, \href
  {https://ui.adsabs.harvard.edu/abs/2012ApJ...747..111W} {747, 111}

\bibitem[\protect\citeauthoryear{{Zhao}, {Gandhi}, {Dashwood Brown}, {Knigge},
  {Charles}, {Maccarone}  \& {Nuchvanichakul}}{{Zhao} et~al.}{2023}]{Zhao23}
{Zhao} Y.,  {Gandhi} P.,  {Dashwood Brown} C.,  {Knigge} C.,  {Charles} P.~A.,
  {Maccarone} T.~J.,   {Nuchvanichakul} P.,  2023, \mn@doi [\mnras]
  {10.1093/mnras/stad2226}, \href
  {https://ui.adsabs.harvard.edu/abs/2023MNRAS.525.1498Z} {525, 1498}

\makeatother
\end{thebibliography}
\bsp	
\label{lastpage}
\end{document}